\documentclass[journal]{IEEEtran}
\ifCLASSINFOpdf
\else
\fi
\usepackage{adjustbox}
\usepackage{cite}
\usepackage{amsmath,amssymb,amsfonts}
\usepackage{amsthm}
\usepackage{caption}
\usepackage{graphicx}
\usepackage{textcomp}
\usepackage{bm}

\newtheorem{theorem}{Theorem}[section]

\newtheorem{assum}{Assumption}[section]
\usepackage{xcolor}
\usepackage[english]{babel}
\usepackage[utf8]{inputenc}
\usepackage{tikz}
\usetikzlibrary{shapes.geometric, arrows}
\usepackage{algorithm2e}
\usepackage{subcaption}
\usepackage{bbding}
\usepackage{pifont}
\usepackage{wasysym}
\usepackage{amssymb}
\usepackage{float}
\usepackage{amsmath}
\usepackage{multicol,lipsum}

\usepackage{algpseudocode}
\usepackage{etoolbox}
\usepackage{rotating}
\usepackage{colortbl}
\usepackage{nomencl}
\makeatletter
\newcommand*{\algrule}[1][\algorithmicindent]{%
  \makebox[#1][l]{%
    \hspace*{.2em}% <------------- This is where the rule starts from
    \vrule height .75\baselineskip depth .25\baselineskip
  }
}

\newcount\ALG@printindent@tempcnta
\def\ALG@printindent{%
    \ifnum \theALG@nested>0% is there anything to print
    \ifx\ALG@text\ALG@x@notext% is this an end group without any text?
    \else
    \unskip
    \ALG@printindent@tempcnta=1
    \loop
    \algrule[\csname ALG@ind@\the\ALG@printindent@tempcnta\endcsname]%
    \advance \ALG@printindent@tempcnta 1
    \ifnum \ALG@printindent@tempcnta<\numexpr\theALG@nested+1\relax
    \repeat
    \fi
    \fi
}
\patchcmd{\ALG@doentity}{\noindent\hskip\ALG@tlm}{\ALG@printindent}{}{\errmessage{failed to patch}}
\patchcmd{\ALG@doentity}{\item[]\nointerlineskip}{}{}{} % no spurious vertical space
\usepackage{xfrac}

\addto\captionsenglish{}
\usepackage{float}
\begin{document}
\title{Data-Driven Dynamic State Estimation of Photovoltaic Systems via Sparse Regression Unscented Kalman Filter}

\author{Elham Jamalinia,~\IEEEmembership{Student Member,~IEEE,}
        Zhongtian Zhang,~\IEEEmembership{Student Member,~IEEE,}
        Javad Khazaei,~\IEEEmembership{Senior Member,~IEEE,}
        Rick S. Blum,~\IEEEmembership{Fellow,~IEEE}
        % <-this % stops a space
\thanks{This research was under support from the National Science Foundation under Grant NSF-EPCN 2221784. E. Jamalinia, Z. Zhang, J. Khazaei, and R. S. Blum are with the Electrical and Computer Engineering department at Lehigh University, PA, USA. (E-mails: $elj320@lehigh.edu$, $zhz819@lehigh.edu$ $jak921@lehigh.edu$, and $rblum@eecs.lehigh.edu$).}% <-this % stops a space
}

\maketitle

\begin{abstract}

Dynamic state estimation (DSE) is vital in modern power systems with numerous inverter-based distributed energy resources including solar and wind, ensuring real-time accuracy for tracking system variables and optimizing grid stability. This paper proposes a data-driven DSE approach designed for photovoltaic (PV) energy conversion systems (single stage and two stage) that are subjected to both process and measurement noise. The proposed framework follows a two-phase methodology encompassing ``data-driven model identification" and ``state-estimation." In the initial model identification phase, state feedback is gathered to elucidate the dynamics of the photovoltaic systems using nonlinear sparse regression technique. 
Following the identification of the PV dynamics, the nonlinear data-driven model will be utilized to estimate the dynamics of the PV system for monitoring and protection purposes. To account for incomplete measurements, inherent uncertainties, and noise, we employ an ``unscented Kalman filter," which facilitates state estimation by processing the noisy output data.
Ultimately, the paper substantiates the efficacy of the proposed sparse regression-based unscented Kalman filter through simulation results, providing a comparative analysis with a physics-based DSE.
\end{abstract}


% \fi
\IEEEpeerreviewmaketitle

\section{Introduction}
\IEEEPARstart{O}{ver} the past few years, there has been a surge in the penetration of power generated by solar photovoltaic (PV) systems into the electricity grid \cite{roman2006intelligent}.
% This uptick can be attributed to technological advancements, resulting in cost reductions for power electronic devices, coupled with various incentive programs introduced by governments to promote sustainable energy sources\cite{khazaei2020small}.
With a high penetration of these inverter-dominated distributed energy resources into the electricity grid, the critical role of dynamic state estimation (DSE) in enhancing the reliability, security analysis, and control of modern power systems has become increasingly apparent \cite{meliopoulos2023dynamic}, \cite{fan2023integrated}. In power systems' DSE, Kalman filtering has been widely studied \cite{zhao2017framework,wang2023resilient,zhao2019power}. However, the original Kalman filter is suited for linear systems, posing challenges in the context of prevalent nonlinear power systems \cite{liu2020comparisons}. To address this limitation, the extended Kalman filter (EKF) was developed, employing model linearization and a first-order Taylor expansion in each time step to accommodate nonlinearities \cite{ghahremani2011dynamic}.  While effective for mild nonlinear systems, EKF falls short in highly nonlinear scenarios. To address this, the unscented Kalman filter (UKF) was developed, estimating states by propagating sigma points via unscented transformation to the next time step and tracking their mean and covariance \cite{valverde2011unscented,qi2016dynamic,ghahremani2011online}. \par
While there has been commendable progress in DSE for power systems, it is important to note that a majority of the efforts have been built upon the assumption of having a precise understanding of the system's exact model\cite{mohammadrezaee2021dynamic, marchi2020loss}. This assumption, however, may not always hold true, particularly within the complex and dynamic nature of power systems with high penetration of renewables such as solar PV systems.
% The UKF,  akin to other Kalman filtering methods, demands precise knowledge of the system's exact model, a requirement that poses challenges in practical dynamical systems \cite{zhao2016robust}. This difficulty is notably pronounced in domains such as power systems, including PV systems.
% \par
In PV systems, nonlinearity of the dynamics, stochastic variations in generation, and model/parameter uncertainties exist in the dynamic model \cite{mokaribolhassan2022distribution}. The uncertainties associated with PV systems extend beyond stochastic variations, encompassing unknown factors and uncertainties arising from sources such as sensor errors \cite{khazaei2020small}.
% In light of the inherent uncertainties associated with PV systems, the imperative nature of model identification becomes pronounced \cite{humada2020modeling,ortiz2005analytical}. 
Recognizing the dynamic and uncertain nature of PV systems motivates our pursuit of a data-driven DSE approach, where the identification of the underlying model becomes critical for an adaptive and accurate state estimation.

% Employing a data-driven model identification approach in state-estimation addresses the current dependence on accurate models in existing model-based DSE techniques. Instead, it utilizes available measurements to reveal inherent model uncertainties, providing a more nuanced comprehension of the dynamic behavior of inverter-dominated power systems. However, constructing accurate models from data remains challenging, given the uncertain nature inherent to PV systems \cite{el2023photovoltaic}. \par

\par
\begin{table*}
\centering
\caption{State-of-the-art study on model identification of power systems}
\label{tab.def1}
\begin{tabular}{c c c c c c}  
 \hline
Method  & Reference & Prior Knowledge-independent & Linearization-independent &   Adaptable for model variation \\ [0.5ex] 
 \hline
DMD & \cite{dmdp1, dmdp2} & \CheckmarkBold & \ding{56} & \ding{56} \\ 
Neural Network & \cite{huang2022applications} & \CheckmarkBold & \CheckmarkBold & \ding{56}\\
Koopman Theory & \cite{budivsic2012applied, mauroy2019koopman} & \ding{56} & \CheckmarkBold & \ding{56} \\ 
Least Squared Parameter Identification & \cite{di2013dynamic, alqahtani2016data} & \ding{56} & \CheckmarkBold & \CheckmarkBold \\ 
Genetic Algorithm &\cite{dizqah2014accurate}& \ding{56} & \CheckmarkBold & \CheckmarkBold  \\ 
Sparse Regression & \cite{syndy,syndy1,brunton2016discovering} & \CheckmarkBold & \CheckmarkBold & \CheckmarkBold \\ 
 [0.1ex] 
 \hline
\end{tabular}
\end{table*}
% \begin{table*}
% \centering
% \caption{State-of-the-art study on dynamic state estimation of power systems}
% \label{tab.def2}
% \begin{tabular}{c c c c c c}  
%  \hline
% Method  & Reference & Physical model-independent & Linear model-independent &   Adaptable for fault\\ [0.5ex] 
%  \hline
% Kalman Filter & \cite{muscas2020new} & \ding{56} & \ding{56} & \ding{56} \\ 
% EKF & \cite{ghahremani2011dynamic,dash1999frequency} & \ding{56} & \CheckmarkBold & \ding{56}\\
% UKF & \cite{zhao2016robust,qi2016dynamic} & \ding{56} & \CheckmarkBold & \CheckmarkBold \\ 
% Sparse Regression UKF & Proposed Method & \CheckmarkBold & \CheckmarkBold & \CheckmarkBold \\ 
%  [0.1ex] 
%  \hline
% \end{tabular}
% \end{table*}
In the context of data-driven model identification of nonlinear dynamics, various studies have explored data-driven methods, including dynamic mode decomposition (DMD) \cite{dmdp1, dmdp2}, sparse identification of nonlinear dynamics (SINDy) \cite{syndy, syndy1, fasel2021sindy}, neural networks \cite{huang2022applications}, and koopman theory \cite{budivsic2012applied, mauroy2019koopman}. Table \ref{tab.def1} illustrates a comparison between various model identification methods found in the literature. Among these techniques, sparse regression exhibits versatility in its application to nonlinear systems and demonstrates adaptability during model variations, requiring minimal prior knowledge of the underlying model \cite{syndy, syndy1}. 
% These methods, while diverse in their approaches, all contribute to the overarching goal of accurately modeling the complex dynamics of cyber-physical systems.
% While DMD handles high-dimensional data, it heavily relies on a linear dynamics assumption. Neural-network-based approaches, on the other hand, demand a substantial amount of training data and are often criticized for their lack of interpretability \cite{syndy, syndy1}. Koopman theory, linking DMD to nonlinear dynamics through an infinite-dimensional linear operator, holds promise but does not guarantee convergence to a finite-dimensional space for many systems \cite{mauroy2019koopman,xu2023data}. The SINDy emerges as a particularly promising avenue by leveraging the sparse regression technique to accurately identify the unknown dynamics of nonlinear systems \cite{syndy, syndy1}. 
One of the major advantages of sparse regression lies in its inherent sparsity, enabling easy implementation, reduced training time, and the formulation of an interpretable model that often outperforms other identification techniques. However, its performance for DSE approaches, specifically for nonlinear PV systems with inherent processes and measurement uncertainties, has not been explored yet.
% Notable advancements in physics-based DSE in power systems \cite{dang2020robust,singh2013decentralized,ghahremani2011dynamic} has been in the literature, which assume a full knowledge of system dynamics is known.
% Table \ref{tab.def2} presents an overview of diverse Kalman filter-based methods, emphasizing the unique advantages of our proposed sparse regression UKF. Noteworthy is our approach's independence from prior model knowledge, its data-driven nature, and its adaptability in the face of model variations.

% there is need for mo research that utilizes data-driven models for DSE purposes. In addition, a \textbf{data-driven} and \textbf{model-free} dynamic state estimation for PV systems remains an unexplored and pivotal area. 

\subsection{Our contribution}
To address this challenge, this paper proposes a data driven sparse regression-based UKF for DSE of PV systems. Our contributions can be summarized as follows:
\begin{figure*}
\centerline{\includegraphics[scale=0.85]{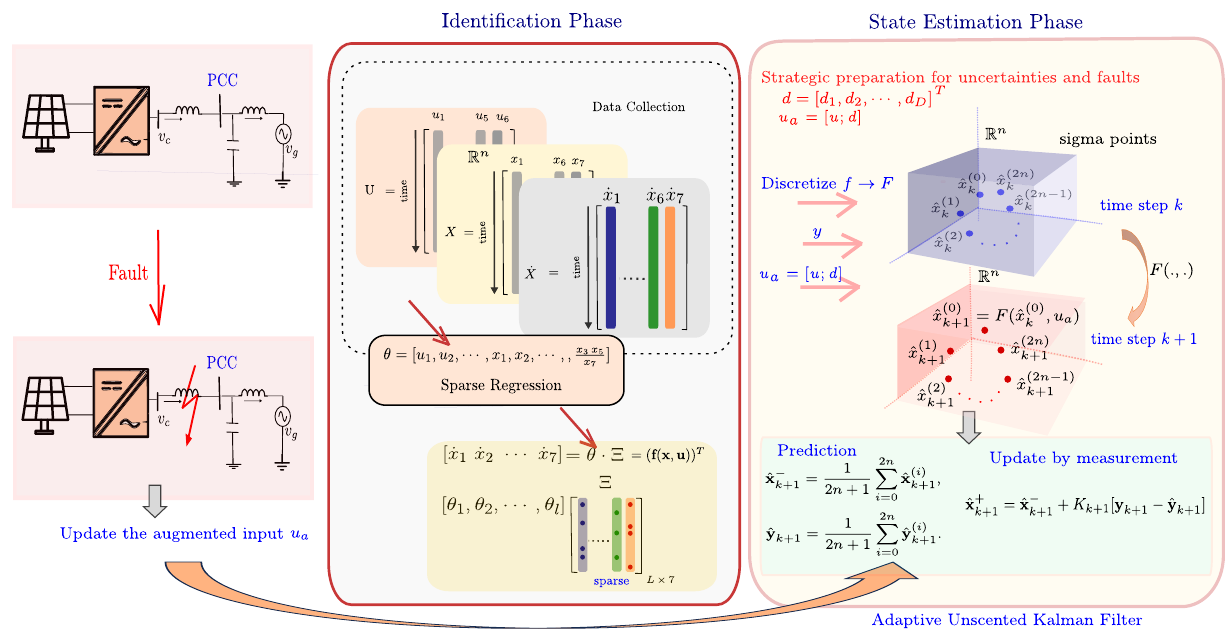}}
\caption{The schematic of sparse regression UKF for data-driven dynamic state estimation of PV systems.}
\label{fig.overallnew}
\end{figure*}
\begin{itemize}

\item  We develop a {model-free}, {data-driven} technique designed for the precise estimation of dynamic states within a solar PV energy conversion system (single stage and two stage). While there have been several works addressing dynamic state estimation in power systems \cite{ghahremani2011online, dang2020robust}, the prevailing approach has predominantly involved model-based methodologies. Diverging from this trend, our work extends the existing research paradigm by explicitly acknowledging the absence of a known model.
\item The concept of data-driven and model-free dynamic state estimation has been previously explored in the literature \cite{surana2016linear, wang2022koopman, netto2018robust}. These studies predominantly center around the Koopman operator, aiming to derive a linear representation of system dynamics followed by the application of the conventional Kalman filter. It is worth noting that the quest for a finite-dimensional subspace using the Koopman operator encounters challenges, particularly in the context of power systems. The intrinsic nonlinearity of power systems renders the task of finding such a subspace not always feasible. In our research, we direct our attention towards sparse regression. This approach, with its accurate nonlinear model identification capability compared to the Koopman operator that lifts the nonlinear dynamics to an infinite dimensional linear space, offers ease of implementation and provides a nonlinear representation of the system\cite{brunton2016discovering}.

% \item We present an innovative contribution through the development of a \textbf{model-free}, \textbf{data-driven} technique designed for the precise and adavptive estimation of states within a PV system.
% \item While prior studies in the identification domain have concentrated on parameter identification for PV systems assuming a known model with unknown parameters \cite{el2023photovoltaic,yu2017parameters,chen2023research}, our research extends this domain by incorporating a sparse regression method. This proposed method enables us to deduce the model of the PV system, even when a pre-existing model is absent.
\item The dynamic model of a PV system connected to the grid is subject to change in real-time (i.e., change in the grid impedance due to faults) \cite{mokaribolhassan2022distribution}.  This temporal variability poses a challenge for conventional DSE approaches that utilize a fixed model or need to calibrate the model using simulations. The evolving nature of the model introduces the need for adaptability. Instead, our proposed approach involves the implementation of an adaptive data-driven model identifier within the UKF that can adaptively update the model parameters in close to real-time only using available measurements and without the need for simulations or model calibration techniques.

% \item This paper introduces an improved sparse regression technique that systematically selects feature values by assessing their relationship with the ratio of the dominant element concerning the sparsity tuning parameter, denoted as $\gamma$. This enhancement aims to provide a more refined and nuanced approach to feature selection within the sparse regression framework.
\end{itemize}

\section{Problem Formulation}
The proposed framework for data-driven DSE of PV systems is depicted in Fig. \ref{fig.overallnew}. The main objectives of a data-driven DSE are twofold: first, to identify the transition function, and second, to estimate the states of the system. The problem of model identification and state estimation for PV systems is addressed for both single-stage and two-stage PV systems. The dynamics of a photovoltaic (PV) system, subject to both additive process and measurement noise, is expressed as
 \begin{align}\label{eq.function}
     \dot{\mathbf{x}}&=\mathbf{f}(\mathbf{x},\mathbf{u})+\mathbf{w},\\
     \mathbf{y}&=\mathbf{g}(\mathbf{x},\mathbf{u})+\mathbf{v},
 \end{align}
In these equations, the system is characterized by $n$ state variables $\mathbf{x}\in\mathbb{R}^n$ and $d$ inputs $\mathbf{u}\in\mathbb{R}^d$. The functions $\mathbf{f}(\mathbf{x},\mathbf{u})$ and $\mathbf{g}(\mathbf{x},\mathbf{u})$ represent the transition and measurement functions, respectively. 
 Furthermore, the system experiences process noise $\mathbf{w}\in \mathbb{R}^{n}$ and measurement noise $\mathbf{v}\in \mathbb{R}^{m}$, both assumed to follow Gaussian distributions, i.e., $\textbf{w}_k\sim N(\textbf{0},\textbf{Q}_k)$, and $ \textbf{v}_k\sim N(\textbf{0},\textbf{R}_k)$, with $\textbf{Q}_k \in \mathbb{R}^{n\times n}$ and $\textbf{R}_k \in \mathbb{R}^{m\times m}$ representing the covariances of process noise and measurement noise, respectively. The transition function, $\mathbf{f}$ in \eqref{eq.function}, is considered unknown. It's worth noting that if the noise statistic is not known, alternative methods can be employed \cite{zhao2016robust}. However, it is important to clarify that the main focus of this work centers around data-driven DSE, and exploring DSE with unknown noise falls outside the primary scope of this paper. This aspect remains a topic for future research.
\begin{figure}
\centerline{\includegraphics[scale=0.85]{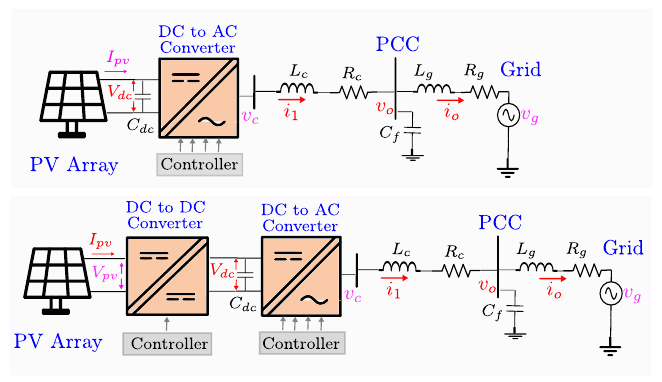}}
\caption{Circuit diagram of single stage and two stage PV systems}
\label{fig.overall1}
\end{figure}
\subsection{Single Stage PV system}
The physical model of the single stage PV system interconnected with the main grid at a point of common coupling (PCC) is expressed in below and is shown in Fig. \ref{fig.overall1}. The system comprises a PV array, a three-phase voltage source converter (VSC), and a low-pass filter. 
\begin{align}
\dot{\textbf{x}}_A&=A\cdot \textbf{x}_A+B\cdot \textbf{u},\label{ACside}\\
    \dot{V}_{dc}&=-\frac{3}{2C_{dc}}\frac{V_{gd}I_{od}}{\,V_{dc}}+\frac{1}{C_{dc}}I_{PV},\label{DClink}
\end{align}
     where
    \begin{align}
      \textbf{x}_A&=[I_{1,d},I_{1,q},I_{{o},d},I_{{o},q},V_{{o},d},V_{{o},q}]^T,\\\nonumber
      \textbf{x}&=[x_A;V_{dc}],\\\nonumber
      \textbf{u}&=[V_{cd},V_{cq},V_{gd},V_{gq},w_0,I_{pv}]^T,
    \end{align}
     % $\textbf{x}_A=[x_1,x_2,x_3,x_4,x_5,x_6]^T$, $\textbf{x}=[x_A;x_7]$ is the state variable of the system, $\textbf{u}=[u_1,u_2,u_3,u_4,u_5,u_6]^T$ $A$ is the input vector and $A$ is
     \begin{align}
         A=&\begin{bmatrix}
         {-\frac{R_c}{L_c}} & w_0 & 0 & 0 & -\frac{1}{L_c} & 0\\
         -w_0 & {-\frac{R_c}{L_c}} & 0 & 0 & 0 & {-\frac{1}{L_c}}\\
         0 & 0 & {-\frac{R_g}{L_g}} & w_0 & \frac{1}{L_g} & 0\\
         0 & 0 & -w_0 & {-\frac{R_g}{L_g}} & 0& {\frac{1}{L_g}}\\
         \frac{1}{C_f} & 0 & -\frac{1}{C_f} & 0 & 0 & w_0\\
         0 &  \frac{1}{C_f} & 0 & -\frac{1}{C_f} & w_0 & 0
     \end{bmatrix},\\
     B=&\begin{bmatrix}
         \frac{1}{L_c} & 0 & 0 & 0 & 0\\
         0 & \frac{1}{L_c} & 0 & 0 & 0\\
          0 & 0 & -\frac{1}{L_g} & 0 & 0\\
          0 & 0 & 0 & -\frac{1}{L_g} & 0\\
          0 & 0 & 0 & 0 & 0\\
          0 & 0 & 0 & 0 & 0
     \end{bmatrix}. \label{ab}
     \end{align}
where the model parameters are adopted from \cite{khazaei2020small}. The
{ac-side dynamics are defined in equation \eqref{ACside}, where $V_{cd}$,$V_{cq}$,$I_{1,d}$ and $I_{1,q}$ are the output voltages and currents of the inverter in dq-frame, $I_{{o},d}$,$I_{{o},q}$,$V_{{o},d}$ and $V_{{o},q}$ are currents and voltages at the PCC in dq-frame, $V_{gd}$ and $V_{gq}$ are the grid voltage components in dq-frame. The dc-side dynamics is expressed as
\begin{align}
    &\frac{1}{2}C_{dc}v_{dc}^2s = P_{PV}-P_g \label{vdc} 
\end{align}
 where $V_{dc}$ represents the DC-link voltage and $I_{pv}$ is the output current of the PV array. Equation \eqref{vdc} can be written in time-domain as \eqref{DClink} \cite{khazaei2020small}.
}

\subsection{Two Stage PV system}

The two-stage PV system shares a similar structure with the single-stage PV system with an addition of a DC/DC converter. The physical model of the two stage PV system, Fig.\ref{fig.overall1}, has $8$ state variables with the state space
\begin{align}
 \dot{\textbf{x}}_A&=A\cdot \textbf{x}_A+B\cdot \textbf{u}\label{ac2}\\
    \dot{I}_{PV} &= \frac{1}{L_{b}}V_{PV}-\frac{(1-d_{ref})}{L_{b}}V_{dc} \label{ipv1}\\
     \dot{V}_{dc} &= \frac{(1-d_{ref})}{C_{dc}}I_{PV}-\frac{1}{C_{dc}}I_{dc} \label{vdc1} 
\end{align}
where $v_{PV}$ and $i_{PV}$ are the output voltage and current of the PV array, $v_{dc}$ is the DC-link voltage, $C_{dc}$ is the value of the capacitor on the DC side, and $d_{ref}$ is
defined as the duty cycle of the DC/DC converter. The input current of the VSC, $i_{dc}$, can be represented by the power balance equation \cite{khazaei2020small}
\begin{align}
    &V_{dc}I_{dc} = \frac{3}{2}(V_{od}I_{od}+V_{oq}I_{oq})\label{balance}\\
    &I_{dc} = \frac{3}{2V_{dc}}(V_{od}i_{od}+V_{oq}V_{oq})\label{idc}
\end{align}

\section{Data-Driven Model Identification}
Assume a dynamical system is modeled as:
\begin{equation}\label{eq.dynamic}
    \dot{\textbf{x}}(t)=f(\textbf{x}(t),\textbf{u}(t))
\end{equation}
where $\textbf{x}$ is the state vector, \textbf{u} is the input vector and $f(\textbf{x},\textbf{u}):\mathbb{R}^n\times \mathbb{R}^d\to \mathbb{R}^n$. The underlying principle of sparse identification is that the function \(f\) consists of only a few active terms. To identify the governing equations of the nonlinear system, we need to collect the values of states, the time derivatives of states and the inputs over $m$ time steps during the identification phase \cite{brunton2016discovering}.

\subsubsection{Data Collection}
Since all measurements contain noise, we need to smooth out the data to increase reliability before identification and in our work, we apply Savitzky-Golay filtering \cite{khazaei2022model}.
Based on the filtered data, we define the following matrices
{\small
\begin{align}\label{eq.x} 
     &\textbf{X}=\begin{bmatrix}
\textbf{x}^T(t_1)\\
\textbf{x}^T(t_2)\\
\vdots\\
\textbf{x}^T(t_m)
\end{bmatrix},\ \dot{\textbf{X}}=\begin{bmatrix}
\dot{\textbf{x}}^T(t_1)\\
\dot{\textbf{x}}^T(t_2)\\
\vdots\\
\dot{\textbf{x}}^T(t_m)
\end{bmatrix},\ \textbf{U}=\begin{bmatrix}
        \textbf{u}^{T}(t_1)\\
        \textbf{u}^{T}(t_2)\\
        \vdots\\
         \textbf{u}^{T}(t_m)
    \end{bmatrix} 
\end{align}}
where $\bm{X}$, $\bm{U}$,  and $\bm{\dot{X}}$ are the matrices of the filtered states, inputs, and time derivative of states, respectively.

\subsubsection{Library of Candidate Functions}
To identify the terms on the right-hand side of (\ref{eq.dynamic}), a library of candidate functions is chosen. This library may encompass constants, polynomials, and trigonometric terms, e.g., $\theta(\bm{X,U}) =\begin{bmatrix}\bm{U}  & \bm{X} & \bm{{P_2}(X,U)} &\bm{{P_3}(X,U)} & \cdots & sin(\bm{X,U})& \cdots\end{bmatrix}$, where $P_2(\bm{X,U})$ given in \eqref{eq.P2} represents the second-order combinations of elements of $\bm{X}$ and $\bm{U}$.
\begin{align}\label{eq.P2}
P_2(\bm{X},\bm{U})=\begin{bmatrix}
        x_1(t)u_1(t)& x_1(t)u_2(t) & \cdots & x_n(t)u_d(t)
    \end{bmatrix}
\end{align}

 After $m$ time steps, the matrix $\bm{\Theta}$, which augments the matrix $\bm{\theta}$ at each time step, is defined as (stacking rows):
{\small
\begin{align}\label{eq.Theta2}
  \Theta(\mathbf{X},\mathbf{U})= \begin{bmatrix}
        \theta(\mathbf{X}(t_1),\mathbf{U}(t_1))\\ 
    \theta(\mathbf{X}(t_2),\mathbf{U}(t_2))\\ 
    \vdots\\
    \theta(\mathbf{X}(t_m),\mathbf{U}(t_m))
   \end{bmatrix}_{m\times L}time\downarrow
\end{align}
} 
% \begin{equation}\label{eq.Library}
% \bm{\Theta}(\textbf{X},\textbf{U})=[\theta_1;\theta_2;\cdots;\theta_m]. 
% \end{equation}
The time derivative of states $\bm{\dot{X}}$ can now be written as a linear expansion of the functions in the library $\bm{\Theta}$
\begin{equation}\label{eq.sparse}
    \dot{\textbf{X}}=\bm{\Theta}(\textbf{X},\textbf{U})\bm{\Xi},
\end{equation}
where $\bm{\Xi}$ is the matrix of coefficients for the candidate functions in $\Theta$ to be obtained.
\subsubsection{Identification by Sparse Regression}
To find $\bm{\Xi}$, a sequentially thresholded least-square optimization problem is solved.
\begin{align}\label{eq.min}
    \xi_h=\arg\min \|\dot{X}_h-\Theta(\mathbf{X},\mathbf{U})\hat{\xi}_h\|_2+\gamma\|\hat{\xi}_h\|_0
\end{align}
where $\xi_h$ is the $h$-th column of $\bm{\Xi}$ represented by $\xi_h=[\xi_1\;\xi_2;\cdots\;\xi_p]^T$, moreover $\|.\|_2$ and $\|.\|$ denote $l_2$ and $l_0$ norms, respectively. Sparsity of the resulting $\bm{\Xi}$ matrix is adjusted with a sparsity promoting hyperparameter $\gamma$, which is usually tuned manually \cite{brunton2016discovering}.
Equation \eqref{eq.min}  which results in a sparse $\bm{\Xi}$, is approximately solved
by the sequential thresholded least squares (STLS) proposed
in \cite{brunton2016discovering} defined as 

\begin{align}\label{eq.xii}
   \hat{ \mathbf{\xi}}_h^0=\mathbf{\Theta}(\mathbf{X},\mathbf{U})^{\dag}\mathbf{X}_h
\end{align}
\begin{align}
    \mathbf{\xi}^{k+1}=\arg\min_{{\hat{\mathbf{\xi}}_h}\in \mathbb{R}^p}\|\mathbf{X}_h-\mathbf{\Theta}(\mathbf{X},\mathbf{U})\hat{\xi}_h\|_2,\;k\geq 0
\end{align}
where $k$ is the iteration number and $\mathbf{\Theta}(\mathbf{X},\mathbf{U})^{\dag}$ is the pseudo-inverse of $\mathbf{\Theta}(\mathbf{X},\mathbf{U})$ and is defined as
\begin{align}
    \mathbf{\Theta}(\mathbf{X},\mathbf{U})^{\dag}:=[\mathbf{\Theta}(\mathbf{X},\mathbf{U})^T\mathbf{\Theta}(\mathbf{X},\mathbf{U})]^{-1}\mathbf{\Theta}(\mathbf{X},\mathbf{U})^T
\end{align}
\subsubsection{Validation of Single Stage and Two Stage PV Models}
In the evaluation of sparse regression applied to a PV system, data pertaining to the state variables is systematically collected in $0.5$ second, with sampling resolution of $10^{-4}$ seconds. 

 We utilized a library of functions including linear terms, polynomials up to degree 3, and sinusoidal functions for identifying PV dynamics.

The sparse matrix $\mathbf{\Xi}$ for single-stage and two-stage PV is shown in Table \ref{table:1} and Table \ref{table:2} for $\gamma=8$ and $\gamma=15$, respectively. 

\begingroup

\setlength{\tabcolsep}{10pt} % Default value: 6pt
\renewcommand{\arraystretch}{1.5} % Default value: 1
\begin{table}
\centering
\caption{Identified $\Xi$ with $\gamma=8$ for single stage PV}
\label{table:1}
\begin{adjustbox}{width=9cm,center}
\begin{tabular}{|c ||c |c |c |c| c| c| c|}  
 \hline
  {$\Xi$}& $\dot{x}_1$ &$\dot{x}_2$ & $\dot{x}_3$& $\dot{x}_4$ & $\dot{x}_5$ & $\dot{x}_6$ & $\dot{x}_7$\\ [0.5ex] 
 \hline
 \hline
$x_1$ & \cellcolor{blue!10}${-133.58}$ & $\cellcolor{blue!50}{-377}$ & $\color{black!25}{2.31\rightarrow}$ $0$ & ${0}$ & \cellcolor{blue!50}${4000}$ & ${0}$ & ${0}$
\\ 
$x_2$&
   \cellcolor{blue!50}{ $377$}
&\cellcolor{blue!10}${-133.58} $& $ {0}$& $\color{black!35}{8.52}\rightarrow $ $0$ & ${0}$ & \cellcolor{blue!50}${4000}$ & ${0}$\\ 
 $x_3$& $0$& $0$& \cellcolor{blue!10}$-133.58$& \cellcolor{blue!50}$-377$ & \cellcolor{blue!10}${-4000}$ & $$\color{black!35}{25.17}$$ & $0$\\ 
$x_4$&$0$&$0$&\cellcolor{blue!50}${377}$&\cellcolor{blue!10}${-133.58}$ & $\color{black!35}{35.8}$ & \cellcolor{blue!10}${-4000}$ & $0$\\ 
 $x_5$& $\color{black!35}{-38.17}$&$0$&$\color{black!35}{38.17}$&${0}$ & $0$ & $\color{black!35}{-377}$ & $0$\\ 
 $x_6$&$ 0$ &$ \color{black!35}{-38.17}$ & $ 0$& $\color{black!35}{38.17}$&$\color{black!35}{377}$ & $ 0$& $0$\\
 $x_7$ &$0$ & $0$&$0$ &$0$&$0$ &$0$ &\color{black!35}$15.31$\\ 
 $u_1$ & $\color{black!35}{38.17}$& $0$ &$0$ &$0$ & $0$ & $0$ & $0$\\
 $u_2$&$0$ & $\color{black!35}{38.17}$& $0$& $0$ & $0$ & $0$ & $0$\\ 
 $u_3$& $0$&$0$& $\color{black!35}{-38.17}$& $0$ & $0$ & $0$ & $0$\\ 
 $u_4$ & $0$&$0$ &$0$ &$\color{black!35}{-38.17}$ & $0$ & $0$ & $0$\\  
 $u_5$ & $0$& $0$&$0$ &$0$ & $0$ & $0$ & $0$\\ 
 $u_6$ &$0$ &$0$ & $0$&$0$ & $0$ & $0$ & \cellcolor{blue!10}$166.66$\\
 $\frac{x_3\,x_5}{\,x_7}$ &$0$ &$0$ & $0$&$0$ &$0$ & $0$&\cellcolor{blue!50}$-250$\\
 [0.1ex] 
 \hline
\end{tabular}
\end{adjustbox}
\end{table}
\endgroup

\begingroup
\setlength{\tabcolsep}{10pt} % Default value: 6pt
\renewcommand{\arraystretch}{1.5} % Default value: 1
\begin{table}
\centering
\caption{Identified $\Xi$ with $\gamma=15$ for single stage PV}
\label{table:2}
\begin{adjustbox}{width=9cm,center}
\begin{tabular}{|c ||c |c |c |c| c| c| c|}  
 \hline
  {$\Xi$}& $\dot{x}_1$ &$\dot{x}_2$ & $\dot{x}_3$& $\dot{x}_4$ & $\dot{x}_5$ & $\dot{x}_6$ & $\dot{x}_7$\\ [0.5ex] 
 \hline
 \hline
$x_1$ & $\cellcolor{blue!15}{-133.58}$ & \cellcolor{blue!50}${-377}$ & $\color{black!15}{2.31}$ & ${0}$ &\cellcolor{blue!50}${4000}$ & ${0}$ & ${0}$
\\ 
$x_2$&\cellcolor{blue!50}$377$&\cellcolor{blue!15}${-133.58} $& $ {0}$& $\color{black!15}{8.52}$ & ${0}$ & \cellcolor{blue!50}${4000}$ & ${0}$\\ 
 $x_3$& $0$& $0$& \cellcolor{blue!15}$-133.58$& \cellcolor{blue!50}$-377$ & \cellcolor{blue!15}$-4000$ & $\color{black!15}{25.17}$ & $0$\\ 
$x_4$&$0$&$0$&\cellcolor{blue!50}${377}$&\cellcolor{blue!15}${-133.58}$ & $\color{black!15}{35.8}$ & \cellcolor{blue!15}$-4000$ & $0$\\ 
 $x_5$&\cellcolor{blue!15} ${-38.17}$&$0$&\cellcolor{blue!15}${38.17}$&${0}$ & $0$ & \cellcolor{blue!15}$-377$ & $0$\\ 
 $x_6$&$ 0$ &\cellcolor{blue!15}$ -38.17$ & $ 0$&\cellcolor{blue!15} $38.17$&\cellcolor{blue!15}$377$ & $ 0$& $0$\\
 $x_7$ &$0$ & $0$&$0$ &$0$&$0$ &$0$ &\color{black!35}$15.31$\\ 
 $u_1$ & \cellcolor{blue!15}$38.17$& $0$ &$0$ &$0$ & $0$ & $0$ & $0$\\
 $u_2$&$0$ & \cellcolor{blue!15}$38.17$& $0$& $0$ & $0$ & $0$ & $0$\\ 
 $u_3$& $0$&$0$& \cellcolor{blue!15}$-38.17$& $0$ & $0$ & $0$ & $0$\\ 
 $u_4$ & $0$&$0$ &$0$ &\cellcolor{blue!15}$-38.17$ & $0$ & $0$ & $0$\\  
 $u_5$ & $0$& $0$&$0$ &$0$ & $0$ & $0$ & $0$\\ 
 $u_6$ &$0$ &$0$ & $0$&$0$ & $0$ & $0$ & \cellcolor{blue!15}$166.66$\\
 $\frac{x_3\,x_5}{\,x_7}$ &$0$ &$0$ & $0$&$0$ &$0$ & $0$&\cellcolor{blue!50}$-250$\\
 [0.1ex] 
 \hline
\end{tabular}
\end{adjustbox}
\end{table}
\endgroup
It is noted that the high-order library terms with zero coefficients are not shown in these two Tables to save space. The identified coefficients closely match the physical model and confirm the accuracy of proposed data-driven technique for PV system identification. 

\section{Adaptive Sparse Regression Unscented Kalman Filter}
\begin{figure}
\centerline{\includegraphics[scale=0.7]{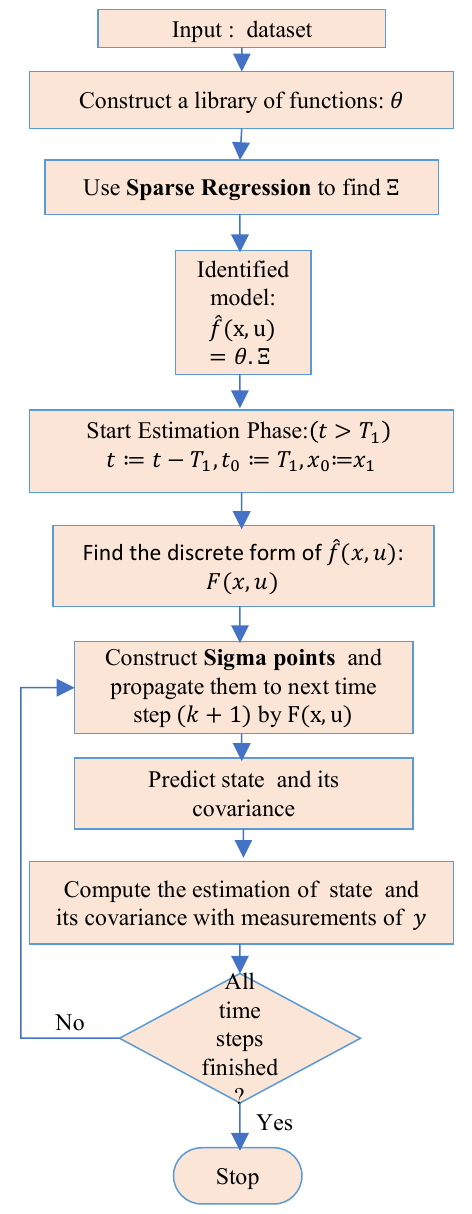}}
\caption{Flowchart of sparse regression UKF for DSE.}
\label{fig.overallMethod}
\end{figure}
In this section, we outline a two-phase methodology devised for addressing model-free dynamic state estimation of PV systems. In the initial identification phase, we introduce a variant of sparse regression designed to address the challenge of determining an optimal sparsity tuning constant, $\gamma$ for effective model identification. Following this, we present a novel adaptation of the unscented Kalman filter, referred to as the adaptive unscented Kalman filter, specifically crafted to accommodate model changes and facilitate state estimation in the presence of such dynamic alterations. The flow chart of the proposed method is shown in Fig.\ref{fig.overallMethod}.
\subsection{Feature Selection Sparse Regression}
Algorithm \ref{alg:featuresparse} introduces an enhanced sparse regression technique that systematically selects feature values by evaluating their relationship to the ratio of the dominant element relative to the sparsity tuning parameter, denoted as $\gamma$. The selection of $\gamma$ is guided by the objective of minimizing the least square error in modeling the training data with the identified model. The algorithm operates as in \textbf{Algorithm \ref{alg:featuresparse}}.
\RestyleAlgo{ruled}
\SetKwComment{Comment}{$\triangleright$ }{ }

\begin{algorithm}
\caption{Feature Selection Sparse Regression}
\label{alg:featuresparse}
\small 
\KwData{$\bm{U}, \bm{X},\dot{\bm{X}}$}
\KwResult{$\bm{\Xi}$}
Select the library of $m$ functions: $\bm{\theta}(\textbf{x},\textbf{u})$\;
Build $\bm{\Theta}(\textbf{X},\textbf{U})$ as in \eqref{eq.Theta2}\;
Find $\bm{\Xi}=\bm{\Theta}(\bm{X},\bm{U})\backslash\dot{\bm{X}}$\;
\For{j=1:n}
{
    $\mu_j\leftarrow \max(\bm{\Xi}(:,j))$\;
    \Comment{comment: $\mu_j$ is the dominant element in $j$-th column.}
    Set $\gamma$\;
    \Comment{comment: $\gamma$ is the sparsity tuning factor.}
    \For{i=1:m}
    {
        \If{$| \bm{\Xi}(i,j)|          < \mu_j/\gamma$}
        {
            $\bm{\Xi}(i,j)=0$\;
            \Comment{Removes the non-dominant elements while retaining the features.}
        }
    }
}
\end{algorithm}
\subsection{Observability}
To ensure accurate estimation of the system's state variables, it is imperative that the system exhibits observability. In the context of linear systems, the verification of observability can be conducted through the application of the Kalman observability theorem \cite{kalman1968lectures}.
\begin{theorem} 
A linear system with state space as $\dot{\textbf{x}}=\mathbf{A}\textbf{x}+\mathbf{B}\textbf{u}$, and measurement $\textbf{y}=\mathbf{C}\textbf{x}$ is observable if and only if the observability matrix $\mathcal{O}$ has full rank.
The observability matrix is defined as:
\begin{align}\label{eq.observ}
    \mathcal{O} = \begin{bmatrix} \mathbf{C} & \mathbf{CA} & \mathbf{CA}^2& \hdots & \mathbf{CA}^{n-1} \end{bmatrix}^T
\end{align} 
\end{theorem}
 In our research, each of the systems—single-stage PV and two-stage PV—can be decomposed into two cascaded subsystems of linear and nonlinear dynamics.  It is imperative to conduct a distinct analysis of the observability of these two cascade subsystems. Specifically, the first subsystem within each system adheres to a linear model characterized by \eqref{ACside} with matrix $A$. Considering $y=\begin{bmatrix}
          x_5&
          x_6
      \end{bmatrix}^T$ will result in a full-rank observability matrix. 

To assess the observability of the second subsystem in a single-stage PV system, it is crucial to recognize that the subsystem comprises a single state variable $x_7$. This underscores the importance of measuring this state for ensuring the observability of the second subsystem and, consequently, the entire PV system. The second subsystem in a two-stage PV system comprises two decoupled state variables, facilitating observability for the entire subsystem by measuring just one of the two state variables. Therefore, choosing $y=\begin{bmatrix}
          x_5&
          x_6&
          x_7
      \end{bmatrix}^T$ will result in observability of both PV systems.

\subsection{Adaptive Unscented Kalman Filter with Augmented Input}
% \subsection{Unscented Kalman Filter}\label{sec.UKF}
With the model identified in the previous section, we employ UKF to estimate the system's states, guided by measurements of the noisy output $\mathbf{y}$. This estimation is carried out under the assumption of having knowledge about the statistics of both the process and measurement noise. Since the UKF is designed for discrete-time models, it is imperative to formulate the discrete-time model before proceeding.

\subsubsection{Discrete Time Model}
Applying the Euler method, we represent the time evolution of states and outputs as

\begin{align}
    \textbf{x}_{k+1}=\textbf{F}(\textbf{x}_k,\textbf{u}_k)+\textbf{w}_k, \quad
    \textbf{y}_k=\textbf{G}(\textbf{x}_k,\textbf{u}_k)+\textbf{v}_k,
\end{align}
where $\bm{F}(.,.)$ and $\bm{G}(.,.)$ are the discrete transition function and measurement function respectively and are defined as
\begin{align}\label{eq.discrete}
    \mathbf{\bm{F}}(\textbf{x}_k,\textbf{u}_k)&= \textbf{x}_{k}+\hat{\textbf{f}}(\textbf{x},\textbf{u},\textbf{v}).\Delta t,\\ 
     \textbf{G}(\textbf{x}_k,\textbf{u}_k)&=\textbf{g}(\textbf{x}_k,\textbf{u}_k).
\end{align}
Within the framework of the UKF, a set of deterministic sample points, known as sigma points, is employed to estimate the mean and covariance of a Gaussian random variable. When these sigma points undergo transformation by a nonlinear function, their true mean and covariance can be accurately calculated up to the 3rd order of the Taylor series expansion \cite{wan2000unscented}.\par
\subsubsection{Prediction Step}
At time step $k$, with the given mean $\bm{\hat{x}^{+}}_{k}$ and covariance $\bm{P}^+_{k}$ of a random variable $\textbf{x}$ in $\mathbb{R}^n$, we calculate $2n+1$ sigma points of the variable, all assigned equal weights, following this procedure\cite{julier2004unscented}
\begin{align}\label{eq.sigma}
 \hat{\textbf{x}}^{(0)}_{k}&=\hat{\textbf{x}}^{+}_{k},\\\hat{\textbf{x}}^{(i)}_{k}&=\hat{\textbf{x}}^{+}_{k}+\Tilde{\textbf{x}}^{(i)},\quad \quad \quad \,\, i=1,\cdots,2n\\
 \Tilde{\textbf{x}}^{(i)}&=\Big(\sqrt{n\bm{P}^+_{k}}\Big)^T_{i},\quad \quad \, \, \, \,\, i=1,\cdots,n\\
\Tilde{\textbf{x}}^{(n+i)}&=-\Big(\sqrt{n\bm{P}^+_{k}}\Big)^T_{i},\quad \quad i=1,\cdots,n
\end{align}
where $\hat{\textbf{x}}^{(i)}_{k}, i=0,\cdots,2n$ is the $(i+1)$-th sigma point of state variable and $\bm{P}^+_{k}$ is the estimated covariance of the state variable at time step $k$. Each sigma point is transformed by the nonlinear function in (\ref{eq.discrete}) to the next time step
\begin{align}
\hat{\textbf{x}}^{(i)}_{k+1}&=\bm{F}(\textbf{x}_k^{(i)},\textbf{u}_k),\\
\hat{\textbf{y}}^{(i)}_{k+1}&=\bm{G}(\textbf{x}_{k+1}^{(i)},\textbf{u}_{k+1}).
\end{align}
Utilizing these propagated sigma points, the mean of both the state and the output at time step $k+1$ are predicted as
\begin{align}\label{eq.pre}
    \hat{\textbf{x}}_{k+1}^{-}=\frac{1}{2n+1}\sum_{i=0}^{2n}\hat{\textbf{x}}^{(i)}_{k+1}, \quad
    \hat{\textbf{y}}_{k+1}=\frac{1}{2n+1}\sum_{i=0}^{2n}\hat{\textbf{y}}^{(i)}_{k+1}.
\end{align}
Consequently, the covariance of the state is predicted by
\begin{equation}\label{eq.cov}
    \bm{P}_{k+1}^{-}=\sfrac{1}{2n}\sum_{i=1}^{2n}(\hat{\textbf{x}}_{k+1}^{(i)}-\hat{\textbf{x}}_{k+1}^{-})(\hat{\textbf{x}}_{k+1}^{(i)}-\hat{\textbf{x}}_{k+1}^{-})^T+\bm{Q}_{k},
\end{equation}
where $\bm{P}_{k+1}^{-}$ is the predicted covariance and $\bm{Q}_{k}$ is added to include the effect of process noise. Furthermore, the prediction of the covariance of the output and the cross-covariance of the state and output at time step $k+1$ is computed as
\begin{align}
    \bm{P}_\textbf{y}&=\dfrac{\sum_{i=0}^{2n}(\hat{\textbf{y}}_{k+1}^{(i)}-\hat{\textbf{y}}_{k+1})(\hat{\textbf{y}}_{k+1}^{(i)}-\hat{\textbf{y}}_{k+1})^T}{2n+1}+\bm{R}_k,\\ \label{eq.covxy}
     \bm{P}_{\textbf{xy}}&=\dfrac{1}{2n+1}\sum_{i=0}^{2n}(\hat{\textbf{x}}_{k+1}^{(i)}-\hat{\textbf{x}}^-_{k+1})(\hat{\textbf{y}}_{k+1}^{(i)}-\hat{\textbf{y}}_{k+1})^T.
\end{align}
\subsubsection{Update Step}
With measurement $\textbf{y}_{k+1}$ and Kalman filter equations \cite{welch1995introduction}, the estimation of the mean and covariance in previous step is updated by
\begin{align}\label{eq.kalman1}
\hat{\textbf{x}}_{k+1}^{+}&=\hat{\textbf{x}}_{k+1}^{-}+K_{k+1}[\textbf{y}_{k+1}-\hat{\textbf{y}}_{k+1}],\\\label{eq.kalman2}
    \bm{P}_{k+1}^{+}&=\bm{P}_{k+1}^{-}-K_{k+1}\bm{P}_\textbf{y}K_{k+1}^T,
\end{align}
where $K_{k+1}$ is the Kalman gain and is calculated by
\begin{equation}\label{eq.kalman3}K_{k+1}=\bm{P}_{\textbf{xy}}\bm{P}_\textbf{y}^{-1}.
\end{equation}
The sigma points for propagation in the subsequent time step are computed based on $\hat{\textbf{x}}^{+}_{k+1}$ and $\bm{P}^{+}_{k+1}$. These two steps, prediction and update, are then reiterated in the process.
\subsubsection{Adaptive UKF}
The system dynamics, accounting for uncertainties in the model and the potential variability of model parameters, are expressed as follows

\begin{align}\label{eq.2function}
\dot{\mathbf{x}} &= \mathbf{f}(\mathbf{x}, \mathbf{u}, \mathbf{d}) + \mathbf{w}, \\
\mathbf{y} &= \mathbf{g}(\mathbf{x}, \mathbf{u}, \mathbf{d}) + \mathbf{v},
\end{align}

Here, $\mathbf{d} = [d_1, d_2, \cdots, d_D]^T$ represents a vector of model parameters susceptible to change and uncertainty.

\begin{assum}
If the model parameter $\textbf{d}$ undergoes a change, such alterations are presumed to be discernible and measurable.
\end{assum}

To enhance the adaptability of UKF, an augmented input is introduced, encompassing both the inputs of the PV system and the parameter vector $\begin{bmatrix}
    u & d
\end{bmatrix}$. 
This augmentation is imperative for dynamically accommodating changes in the system parameters. The updated dynamics of the system, now considering the augmented input, are given by

\begin{align}\label{eq.3function}
\dot{\mathbf{x}} &= \mathbf{f}(\mathbf{x}, \mathbf{u_a}) + \mathbf{w}, \\
\mathbf{y} &= \mathbf{g}(\mathbf{x}, \mathbf{u_a}) + \mathbf{v},
\end{align}

Incorporating this augmented input, the prediction and update steps of the UKF, as defined in Section C.2, must be implemented with the new input definitions. This adaptation ensures that the filtering process can effectively respond to changes in the model parameters, providing accurate and responsive state estimation in dynamic environments.

\subsection{Data-Driven unscented Kalman filter implementation}\label{sec.est}
Following identifying the model of the PV system by Algorithm \ref{alg:featuresparse}, the noisy output $y$ of the generator is observed and the states are estimated by implementing the proposed data-driven UKF via the following steps
\begin{enumerate}
    \item Start the identification phase,
    \item Excite the inputs of the  generator and collect data set for $\textbf{x}$ and $\dot{\textbf{x}}$ and $\textbf{u}$ in each sampling time,
    \item Improve the reliability of the data by a  Savitzky-Golay filter, 
    \item Construct matrices $\bm{X}$, $\dot{\bm{X}}$ and $\bm{U}$ in the form of (\ref{eq.x}),
\item Start Algorithm \ref{alg:featuresparse} to find a sparse $\bm{\Xi}$ ,
\item Write the state space of model as  $    \dot{\textbf{x}}^T=\bm{\theta}(\textbf{x},\textbf{u})\bm{\Xi}$,
% \item The identification phase is finished and start the estimation phase:
% \item Measure output $y$ of the system,
% \item Have an assumption on initial expected value and covariance of the system,
\item Construct sigma points as in (\ref{eq.sigma}) and propagate them to the next time step by the identified model,
\item Predict the mean and covariance of $\textbf{x}$ and cross covariance between $\textbf{x}$ and $\textbf{y}$ by (\ref{eq.pre})-(\ref{eq.covxy}),
\item Update the estimation by Kalman filter equations in (\ref{eq.kalman1})-(\ref{eq.kalman3}).
\end{enumerate}
% \begin{figure}
% \centerline{\includegraphics[scale=0.85]{AlgorithmEdit.pdf}}
% \caption{\footnotesize Schematic of the algorithm for density-guided sparse regression unscented Kalman filter.}
% \label{fig.algorithm}
% \vspace{-0.3in}
% \end{figure}
The schematic of the proposed method is presented in Fig. \ref{fig.overallMethod}.
% The excitation of the system plays a crucial role in gathering sufficient data to accurately capture the dynamics of the system and drive the underlying governing equations. Moreover, it is important that the identification phase occurs during the transient state of the system dynamics when the system is exhibiting dynamic behavior. 
\section{Case Studies}
\subsection{Simulation Settings}
The simulations are executed on a host computer equipped with an {Intel(R) Xeon(R) Silver $4214$R CPU @ $2.40$ GHz} and {$64$ GB RAM}. To accurately emulate the dynamic behavior of PV systems through simulation, a sample time of $0.1$ milliseconds ($ms$) has been established. This choice is grounded in the necessity for precision in implementing UKF algorithms. In practical scenarios, however, the maximum measurement rate of devices such as PMUs is limited to $0.01$ seconds ($s$). To align with the desired sampling time for UKF accuracy, it becomes imperative to employ linear interpolation techniques for real-time measurements, bridging the temporal gap between the actual measurement increments and the specified sample time \cite{zhao2016robust}. 
\begin{figure}
\centerline{\includegraphics[scale=0.95]{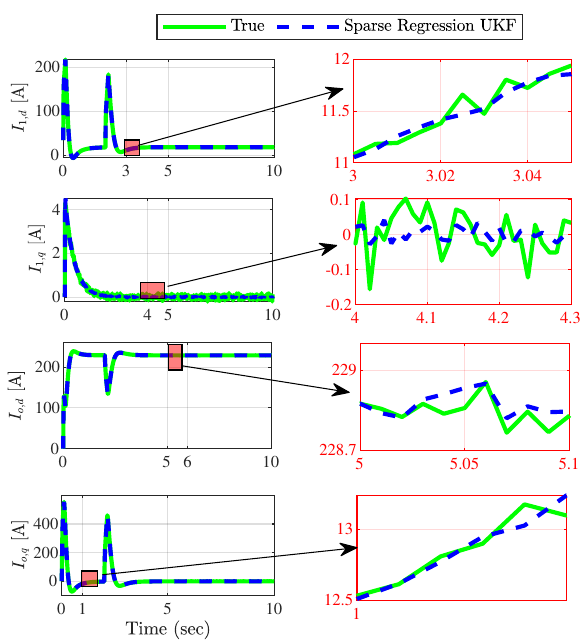}}
\caption{Sparse Regression UKF for single stage PV.}
\label{fig.overall4}
\end{figure}
% \begin{figure}
% \centerline{\includegraphics[scale=0.95]{Figures/TrueTwo.pdf}}
% \caption{Sparse Regression UKF for two stage PV.}
% \label{fig.overall4}
% \end{figure}
% \begin{figure}
% \centerline{\includegraphics[scale=0.95]{Figures/14NovTrueTwo.pdf}}
% \caption{ Sparse regression UKF for two stage PV system.}
% \label{fig.overall}
% \end{figure}

\begingroup

\setlength{\tabcolsep}{10pt} % Default value: 6pt
\renewcommand{\arraystretch}{1.5} % Default value: 1

Following the completion of the model identification process, data-driven UKF with noisy measurements with Gaussian distribution with mean $[0,0,0]^T$ and covariance $10^{-2}\mathbb{I}_{3\times 3}$, was developed as illustrated in Fig.\ref{fig.overall4}. The proposed data-driven DSE can efficiently estimate the underlying dynamics of system states without relying on complex physics-based models. This immediate tracking proficiency enhances the proposed data-driven DSE's real-time adaptability for large-scale power grids where modeling becomes very challenging and contributes to the overall reliability of the DSE technique. In addition, the UKF's inherent ability to assimilate noisy measurements and swiftly converge towards the true states underscores its effectiveness as a robust and responsive estimation technique post-model identification.
\par

\subsection{Effect of the sparsity tuning}
To assess the efficacy of the proposed sparse regression DSE technique across various sparsity tuning parameters ($\gamma$), we considered a two-stage PV system. The sparsity tuning parameter was varied within the set $\gamma=[8, 15, 20, 25, 30]$. In the context of the proposed algorithm, an increase in the sparsity tuning parameter leads to a proportional rise in the number of nonzero elements within the matrix $\Xi$. Conversely, a decrease in the sparsity tuning parameter results in an augmented sparsity of the matrix. Upon examination of the absolute normalized error depicted in Fig.\ref{fig.gamma}, it is evident that the values of $8$ or $10$ yield optimal results for DSE. A sparsity tuning parameter of $\gamma=5$ is found to be misleading, as it leads to the removal of crucial elements from the model, resulting in a significant increase in error of DSE.
\begin{figure}
\centerline{\includegraphics[scale=0.95]{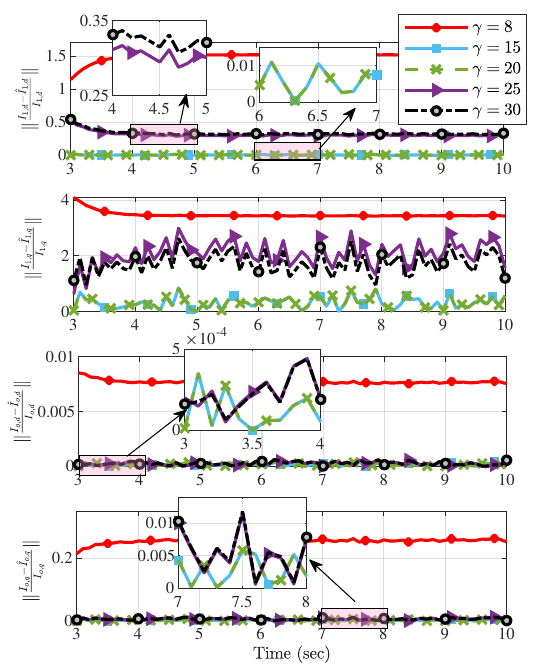}}
\caption{ Effect of $\gamma$ on DSE normalized error via sparse regression UKF in two stage PV}
\label{fig.gamma}
\end{figure}
\subsection{Effect of noise covariance}
In assessing the efficacy of the proposed data-driven DSE in the presence of process noise, independent process noise with a covariance matrix structure of $\sigma \times \mathbb{I}_{7\times 7}$ has been introduced. Here, $\sigma$ assumes values of $0.001$, $0.01$, $0.1$, and $1$. These diverse magnitudes of $\sigma$ are systematically applied to the single-stage PV system, allowing for a comprehensive examination of the sparse regression UKF's robustness under different noise scenarios.  As illustrated in Fig.\ref{fig.noise}, increasing noise intensity reduces the accuracy of the proposed data-driven DSE technique, however, the normalized error reaches to a maximum of 0.004 in the worst case scenario. This small error indicates the robustness of the proposed data-driven DSE under various process noise scenarios. 
\begin{figure}
\centerline{\includegraphics[scale=0.9]{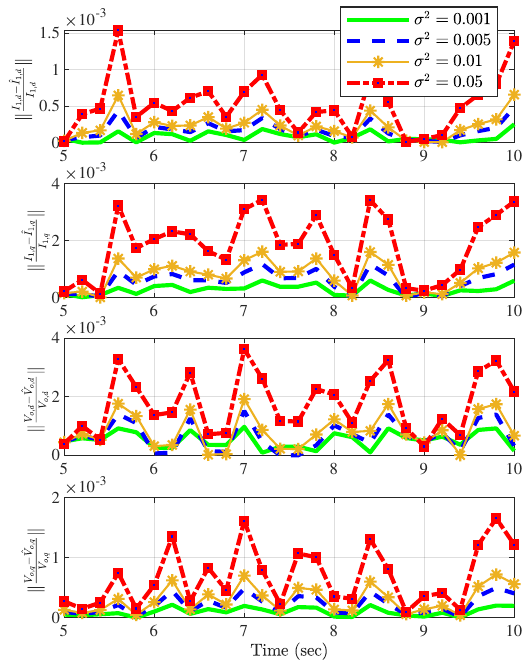}}
\caption{ The effect of covariance of process noise on normalized error of DSE result of sparse regression UKF.}
\label{fig.noise}
\end{figure}

\subsection{Effect of parameter variation}
This case study delves into the repercussions of time-varying parameters on the PV systems. As observed in the dynamic model of single-stage and two-stage PV system, the grid impedance is one of the model parameters and can significantly impact the state-estimation accuracy. Given that the grid impedance can change in real-time (due to topology changes or faults), it is important to test the adaptability of the proposed data-driven DSE when the grid impedance changes. Therefore, at the 20-second mark, an alteration is introduced in the grid resistance $R_g$, transitioning from $3.5$ to $2.90$ $\Omega$. 
% \begin{equation*}
% R_g=3.5\quad \xrightarrow{t=20 s}\quad R_g =2.90
% \end{equation*}
This sudden modification in grid resistance induces an immediate impact on the state variables of the PV system. The comparative analysis between the physics-based DSE and the proposed data-driven DSE is presented in Fig. \ref{fig.ParamTwo} for two-stage PV system. As observed, physics-based DSE exhibits limitations in tracking the true values of the state variables unless a model calibration/update is carried out. In contrast, the adaptive nature of the proposed model identification techniques allows for close-to-real-time update of the model parameters due to a change. This process, encompassing data collection, model retraining, and model updating, was carried out within a 0.5-second interval (from $20$ seconds to $20.5$ seconds), allowing the proposed data-driven DSE to seamlessly continue with the updated model.
The physics-based UKF registers a normalized estimation error of  $20.47$ ,  rendering its estimations unreliable due to the unaccounted parameter change. Conversely, the proposed DSE achieves a commendable normalized estimation error of $0.15$, affirming its efficacy in accurately estimating the true values of the state variables even in the presence of sudden parameter variations. 
\begin{figure}
\centerline{\includegraphics[scale=0.95]{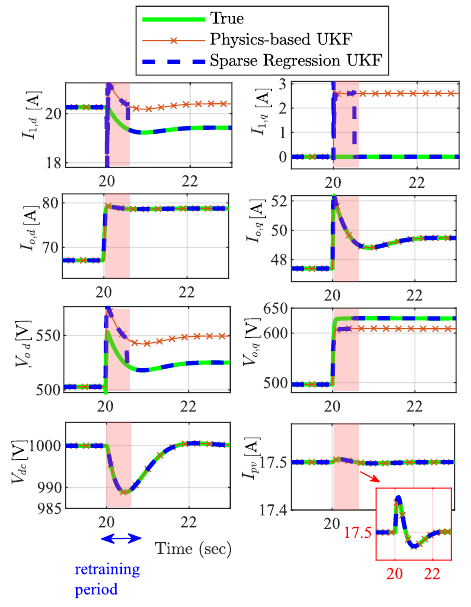}}
\caption{ Proposed data-driven DSE can dynamically update the model online and offer precise real-time estimations of two stage PV system.}
\label{fig.ParamTwo}
\end{figure}

\begin{figure}
\centerline{\includegraphics[scale=0.865]{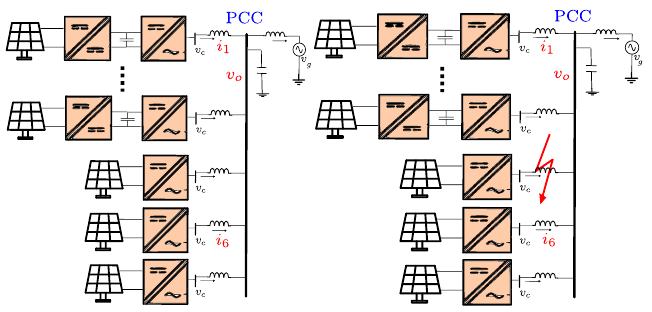}}
\caption{Solar microgrid testbed for system-level validation of the proposed approach.}
\label{fig.Farm}
\end{figure}
\subsection{System-level Validation of the Data-driven DSE}
To assess the efficacy of the sparse regression UKF in a PV microgrid, we examine a microgrid configuration featuring seven PV systems connected according to the topology illustrated in Fig.\ref{fig.Farm}. This microgrid comprises a combination of three single-stage PV systems and four two-stage PV systems. The analysis conducted on this microgrid aims to gauge the proposed data-driven DSE's ability to effectively estimate and track the state variables of each PV system in the presence of dynamic conditions and changing parameters. This evaluation considers the potential system-level advantages of the proposed DSE approach in enhancing estimation accuracy and adaptability across the diverse set of PV systems present in the smart grid.

\begin{figure}
\centerline{\includegraphics[scale=0.8]{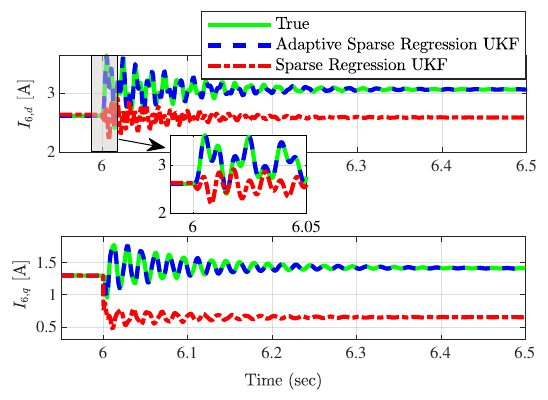}}
\caption{ The effect of system failure in a microgrid with the proposed data-driven DSE.}
\label{fig.overall}
\end{figure}
\begin{figure}
\centerline{\includegraphics[scale=0.8]{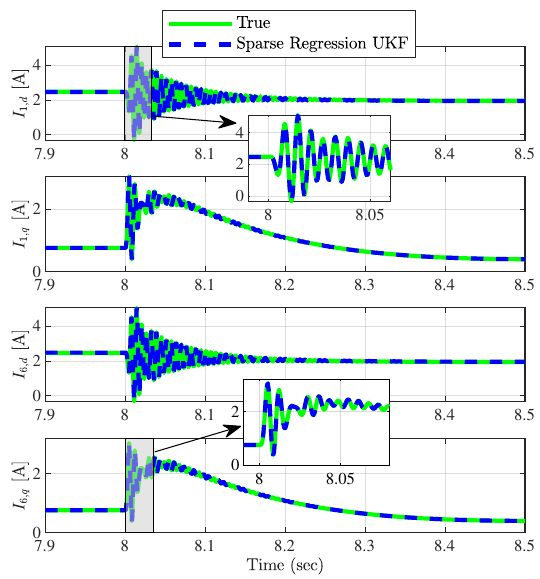}}
\caption{The effect of under voltage fault in a microgrid on performance of proposed data-driven DSE.}
\label{fig.undervoltage}
\end{figure}
\begin{figure}
\centerline{\includegraphics[scale=0.8]{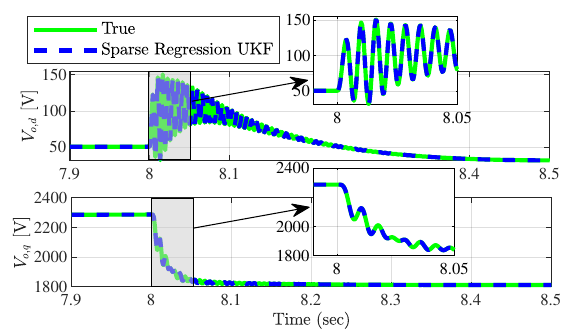}}
\caption{The effect of under voltage fault in a microgrid on performance of DSE result of sparse regression UKF.}
\label{fig.undervoltage2}
\end{figure}
\subsubsection{Equipment failure}
In the initial scenario, the analysis focuses on a failure within the first PV system and the DSE results are depicted in Fig. \ref{fig.Farm}. This failure scenario involves the disconnection of the first PV system, and the subsequent impact on the overall performance and dynamics of the PV microgrid. This failure event introduces a significant perturbation to the microgrid, necessitating a robust estimation and control strategy to manage the ensuing changes in the system behavior. As shown in the figure, the proposed data-driven DSE is effective in accurately tracking the state variables in real-time. The assessment showcases the ability of the proposed DSE approach in providing a reliable state estimation for the remaining PV systems within the microgrid. This scenario serves as a critical test for the proposed DSE approach, addressing challenges posed by system disruptions in a real-world microgrid environment.
\subsubsection{Under voltage fault}
This case study examines the efficacy of the proposed Sparse Regression UKF in response to a significant voltage drop of $20\%$  within the grid, transitioning from 800V to 640V at the 8-second mark. The observed DSE results for the currents in both the first and sixth PV systems are presented in Fig.\ref{fig.undervoltage}. This observation underscore the capability of the proposed data-driven DSE to swiftly and accurately trace the genuine values of the state variables governing the microgrid. This instantaneous tracking ability is crucial for ensuring the system's stability and resilience in the face of dynamic voltage fluctuations.

\section{Conclusion}
In this paper, we delve into the investigation of a data-driven dynamic state-estimation technique that removes the heavy reliance of existing model-based dynamic state estimators to a known physics-based model. Our approach employs a two-phase strategy, encompassing model identification and state estimation, to effectively tackle the challenges typically encountered in conventional model-based dynamic state estimation techniques. The novelty of our proposed method lies in its data-driven modeling approach, which leverages the available measurements to accurately identify the dynamics of PV systems. Subsequently, this data-driven model is harnessed for dynamic state estimation in PV systems, making use of the unscented Kalman filtering technique. The outcomes of our research illustrate the method's proficiency in precisely estimating the states of PV system while being adaptable to real-time model and parameter changes as well as system-level validation for large-scale smart grid applications. Notably, our proposed approach exhibits resilience when faced with PV model uncertainties or system-level faults. These findings provide a robust foundation for future exploration, with potential extensions that can accommodate higher unknown noise within the proposed data-driven dynamic state-estimation framework.

% or

% \subsection or using \label (% by itself


\ifCLASSOPTIONcaptionsoff
  \newpage
\fi

\bibliography{references.bib}
\bibliographystyle{IEEEtran}

\end{document}